\documentclass[final,3p,times,twocolumn,sort&compress]{elsarticle}

\usepackage[inkscapelatex=false]{svg}
\usepackage{amsfonts}
\usepackage{layout} 
\usepackage{amsmath}
\usepackage{amsbsy}
\usepackage{amssymb}
\usepackage{upgreek}
\usepackage{eqnarray}
\usepackage{array}
\usepackage{color}
\usepackage[breaklinks=true,colorlinks=true,linkcolor=black,urlcolor=black,citecolor=blue]{hyperref}  
\usepackage{graphicx}
\usepackage{dcolumn}
\usepackage{bm}
\usepackage{array}
\usepackage{float}
\usepackage{adjustbox}
\usepackage{multirow}
\usepackage{blindtext}
\usepackage{enumitem}
\usepackage{url}
\usepackage{comment}
\usepackage[normalem]{ulem}
\usepackage{xcolor}
\usepackage{listings}
\usepackage{siunitx}

\newcommand{\beq}{\begin{equation}}
\newcommand{\eeq}{\end{equation}}
\newcommand{\bea}{\begin{eqnarray}}
\newcommand{\eea}{\end{eqnarray}}

\newcommand{\descriptor}{Neighbors Map}
\usepackage{placeins}

\let\today\relax
\makeatletter
\def\ps@pprintTitle{%
    \let\@oddhead\@empty
    \let\@evenhead\@empty
    \def\@oddfoot{
         }%
    \let\@evenfoot\@oddfoot
    }
\makeatother

\definecolor{asparagus}{RGB}{50,130,30}
\begin{document}

\title{Neighbors Map: an Efficient Atomic Descriptor for Structural Analysis}

\author[paris]{Arnaud Allera\corref{corresponding}}
\ead{arnaud.allera@cea.fr}
\author[paris]{Alexandra M. Goryaeva}
\author[dam]{Paul Lafourcade}
\author[dam]{Jean-Bernard Maillet}
\author[paris]{Mihai-Cosmin Marinica\corref{corresponding}}
\cortext[corresponding]{Corresponding author}
\ead{mihai-cosmin.marinica@cea.fr}

\address[paris]{Université Paris-Saclay, CEA, Service de recherche en Corrosion et Comportement des Matériaux, SRMP, F-91191, Gif-sur-Yvette, France}
\address[des]{DES–Service de Recherches M\'{e}tallurgiques Appliqu\'{e}es, CEA, Universit\'e Paris-Saclay, F-91191 Gif-sur-Yvette, France}
\address[dam]{CEA, DAM, DIF, F-91297 Arpajon, France; Université Paris-Saclay, LMCE, F-91680 Bruyères-le-Châtel, France}

\date{\today}

\begin{abstract}
Accurate structural analysis is essential to gain physical knowledge and understanding of atomic-scale processes in materials from atomistic simulations.
However, traditional analysis methods often reach their limits when applied to crystalline systems with thermal fluctuations, defect-induced distortions, partial vitrification, etc.
In order to enhance the means of structural analysis, we present a novel descriptor for encoding atomic environments into 2D images, based on a pixelated representation of graph-like architecture with weighted edge connections of neighboring atoms.
This descriptor is well adapted for Convolutional Neural Networks and enables accurate structural analysis at a low computational cost. %
In this paper, we showcase a series of applications, including the classification of crystalline structures in distorted systems, 
tracking phase transformations up to the melting temperature, 
and analyzing liquid-to-amorphous transitions in pure metals and alloys.
This work provides the foundation for 
robust and efficient structural analysis in materials science, opening up new possibilities for studying complex structural processes, which can not be described with traditional approaches. 
\end{abstract}

\maketitle

\section{Introduction}

Structural analysis methods form an essential set of tools to effectively identify and characterize atomic-scale processes that impact materials properties, such as phase transitions \cite{SVM-glasses_Nature_2016,Fan2021,Fukuya2020,Sun2022}, defects formation \cite{ Marinica2012, goryaeva_compact_2023}  and mobility \cite{Marinica2011, lapointe_machine_2022, JacopoAE2022,Cubuk_GBs_PNAS2018},  precipitate growth \cite{Nastar_thermodynamic_2021}, etc.. 
Traditionally, crystal structures are analyzed at the atomic scale based on their similarity to an ideal lattice structure. 
This approach is conceptually simple and, therefore, commonly used for the analysis of Molecular Dynamics (MD) simulations. 
However, it can fail in complex cases, where the structure is significantly distorted due to the presence of defects, intense thermal fluctuations, or large deformations.
Experimentally obtained atomic coordinates, e.g., reconstructed from atom probe tomography (APT), represent one of the most complex cases for analysis because of the relatively low detection rate and low accuracy on atomic positions that are intrinsic to the method.
Another challenge arises from large-scale molecular dynamics simulations as they approach the exa-scale, where efficient on-the-fly analysis methods are needed, to process the vast amount of raw data that is generated, which is too large to be fully stored for later analysis~\cite{Zepeda2017probing,nguyen2021}.

Traditional methods are often based on \textit{ad hoc} criteria established on field expertise to identify that atoms belong to a specific structure type. 
These criteria include local density, centrosymmetry \cite{CSP}, occupancy in Voronoi cells (Wigner-Seitz) \cite{Voronoi-analysis}, detection of Burgers circuits (DXA) \cite{Stukowski_dislo_2012}, and similarity to reference crystal structures, e.g., using polyhedral template matching (PTM) \cite{PTM} and Adaptive Common Neighbor Analysis (a-CNA) \cite{Stukowski_2012}, or bond order parameters \cite{BAA}, among others. 
These geometry-based methods are computationally inexpensive and routinely used for the analysis of MD simulations. 
Most of them are available in widely-used software, such as OVITO \cite{ovito}, that enables fast and convenient visualization and analysis of atomic structures.   However, the common shortcoming of traditional methods is their sensitivity to atomic perturbations \cite{Stukowski_2012, Voronoi-analysis,Polak2022}. Therefore, when applied to complex systems with a high degree of noise or local strain, these methods often require further adaptation or additional processing (e.g., time averaging of atomic positions), which do not always provide a solution.

More recent approaches \cite{Geiger2013, Dietz2017, Reinhart2017, Leitherer2021, Ziletti2018, Goryaeva2020,freitas2020uncovering} tend to compare atomic environments to a statistical distribution of crystalline or defective structures that is learned within a feature space spawned by atomic descriptors, in a way inspired by methods developed in the machine-learning force-field community~\cite{behler2007,bartok2013representing}. 
Interestingly, these methods treat defects, pristine crystalline structures, and non-crystalline systems in a conceptually similar manner, with minimal assumptions about the distribution of the data.
Geiger and Dellago~\cite{Geiger2013} demonstrated the potential of neural networks (NN) trained on a database of descriptors~\cite{behler2007} to efficiently classify crystal structures~\cite{Dietz2017}.
Instead of using NNs, Goryaeva \textit{et al.} \cite{Goryaeva2020, goryaeva_compact_2023} employed statistical distances, within the feature space of a spectral descriptor, bispectrum SO(4) (BSO4) \cite{bartok2013representing}. This approach introduces the notion of distortion score for atoms, which intriguingly exhibits a correlation with the local atomic energy and offers a physically-informed criterion for matching atoms to a reference structure.
Leitherer \textit{et al.}~\cite{Leitherer2021} used a pre-trained Bayesian NN in conjunction with a high-dimensional descriptor (SOAP \cite{bartok2013representing}), where the Bayesian nature of the NN allows for the estimation of uncertainty.
Ziletti \textit{et al.} \cite{Ziletti2018} proposed a method that represented crystals through the calculation of diffraction images, followed by the construction of a deep learning neural network model for classification. 
The recent work by Chung \textit{et al.} \cite{Chung_DC3_2022} follows a similar framework, where the input structure is embedded into the feature space using Steinhardt features \cite{Steinhardt1983}. 
These features, a simplified version of spectral descriptors like BSO4 or SOAP, are then fed into a dense neural network classifier. 
The workflow, named DC3 \cite{Chung_DC3_2022}, improves drastically the accuracy of classification in physical processes such as solid – liquid phase transition for many crystallographic systems.  
Recently, the use of Graph Neural Networks (GNNs) for embedding local environments has gained significant popularity in the fields of chemistry and biology and in the field of materials science, particularly for surrogate models \cite{Fung2021, Reiser2022, Banik2023, Fan2021, Fukuya2020}.

The aforementioned methods always follow a common framework, which can be divided into two main steps: (i) representing atomic data in high dimensions using a descriptor function \cite{bartok2013representing, bartok_machine_2017, deringer2019machine}, and (ii) employing a machine learning (ML) or deep learning (DL) classifier model to process the high-dimensional representation of atomic coordinates.
These state-of-the-art methods are very promising, however, they are often difficult to apply %
in practice, and can be computationally demanding due to their reliance on descriptor calculations in step (i), where a tradeoff between the accuracy of local atomic environment (LAE) representation and computational efficiency \cite{goryaeva2019towards} is commonly required. 
This can be critical in cases where no powerful computation hardware is available, or in HPC applications where analysis can become a bottleneck~\cite{Zepeda2017probing}.
Consequently, there is a pressing need for a lightweight, data-driven approach that can efficiently and accurately characterize atomic positions in various systems, from simple to highly complex cases, with minimal computational cost. 

\begin{figure*}[ht]
    \centering
    \includegraphics[width=\textwidth]{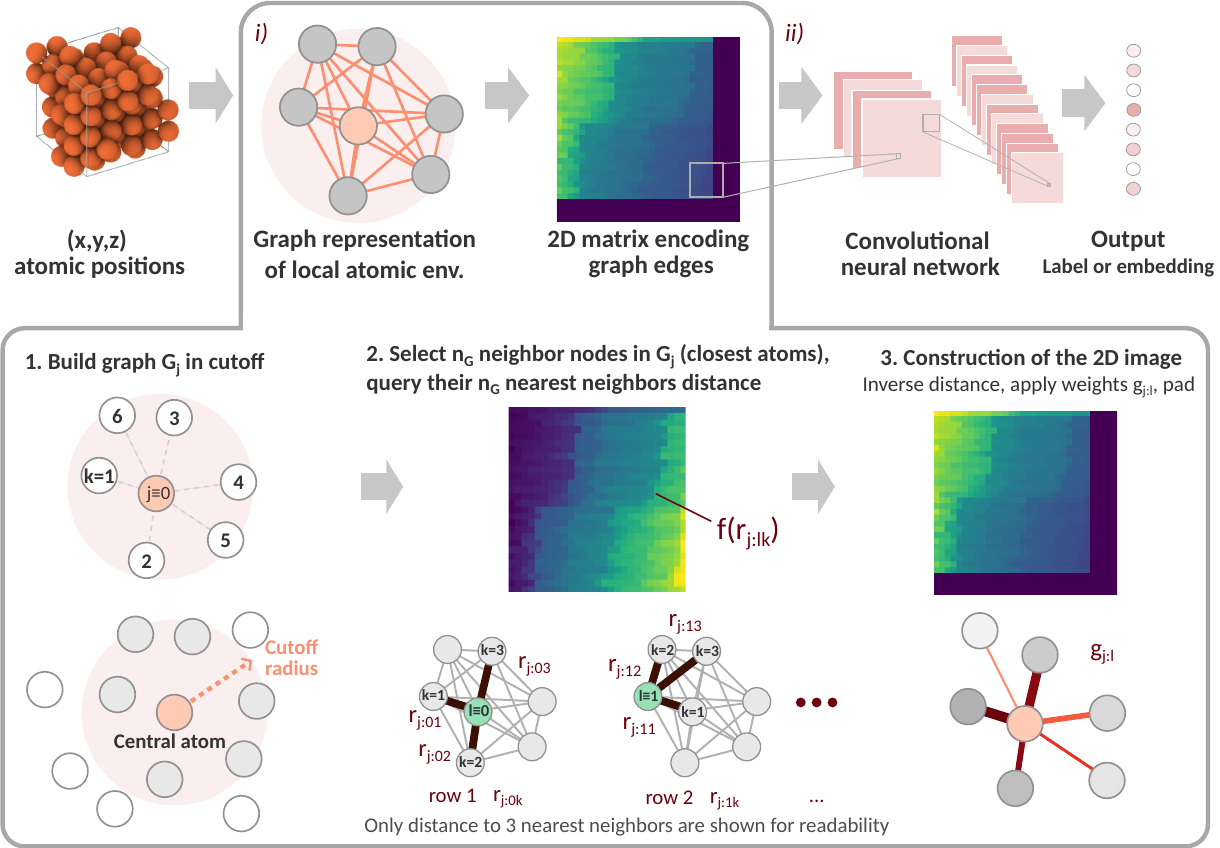}
    \caption{\textbf{Workflow to extract crystallographic environment information from (x,y,z) atomic positions.}
    The procedure includes two main stages.
    Stage \textit{(i)} is depicted within the grey frame and explains the construction of \descriptor{}. Steps 1, 2, and 3 in the lower panel correspond to the procedures described in Sec.~\ref{sec:descriptor_def1}, \ref{sec:descriptor_def2} and \ref{sec:descriptor_def3}, respectively.  
    Stage \textit{(ii)} is the extraction of features in 2D images using the CNN, described in Sec. \ref{sec:cnn}.
    Note that the case $n_G > n(j)$ is represented here for illustration, i.e. the image is padded with a constant (here set to zero) on the right and bottom to match a size of $32\times 32$ pixels.
    }
    \label{fig:f1}
\end{figure*}

In this paper, we present a novel methodology that follows the same framework, associating a descriptor and a ML/DL classification model. 
Illustrated in Fig.~\ref{fig:f1}, our approach combines two key components: 
(i) a lightweight, computationally efficient, invariant descriptor that represents any LAE as a fixed-size image with multiple channels called \descriptor{}, and 
(ii) a DL classification algorithm, with intermediate complexity based on convolutional neural networks to analyze these image representations. 
We address the potential limitations of the descriptor's quality by harnessing the complexity of the neural network, resulting in an efficient and stable workflow capable of accurately identifying atomic environments across a wide range of systems, from simple to highly complex cases.
The descriptor is designed to remain fast to compute, making it suitable for both on-device analysis on a typical workstation, and on-the-fly feature extraction on massively parallel systems, to achieve \textit{in situ} analysis of massive data flow produced by exa-scale simulations. 

After a detailed description of our approach, we present a series of applications that illustrate its robustness, accuracy, and versatility. 
These applications range from classifying common crystalline structures in distorted systems to tracking phase transformations up to the melting temperature, as well as studying the vitrification process in different materials.

\section{Methods}

\subsection{Overview of existing methods for high-dimensional encoding of local atomic environments (LAEs)}

Motivated by the rapid progress of data-driven force field methods, notable strides have been achieved in the realm of descriptor development.
Most commonly, atomic descriptors encode the local geometry of neighboring atoms using either distances and/or angles between atoms \cite{behler2007, rdf_schutt_how_2014, Seko_2014,bartok2013representing}, spectral analysis in spherical harmonics basis of LAEs \cite{bartok_thesis, bartok2013representing},
or based on the scaling wavelets transformation \cite{Eickenberg_Mallat_2017, Hirn_Mallat_2016}. 
Based on the work of Shapeev on the tensorial description of atomic coordinates \cite{Shapeev_MTP,Shapeev_MTP2017}, and on the concept of similarity between pairs of atomic environments known as Smooth Overlap of Atomic Positions (SOAP) \cite{bartok2013representing},  systematic basis were introduced, which preserve permutational and rotational symmetries by writing the total energy as a sum of atomic body-ordered terms. 
By employing an invariant basis constructed from atomic body-ordered polynomials, permutation-invariant polynomials were discovered~\cite{Ortner_allen_atomic_2021, Ortner_oord_regularised_2020}. 
Furthermore, through the use of spectral decomposition and a combination of radial and spherical harmonic functions, it becomes possible to introduce the Atomic Cluster Expansion (ACE) \cite{drautz_atomic_2019, drautz_atomic_2020, lysogorskiy_PACE_2021, dusson_ace_2022},  a  systematic and possibly complete description of the local atomic environment. 
Overall, the capability of the descriptor to map LAEs into the feature space has a direct influence on the quality of embedded force fields and the classification of atomic neighborhoods. 
However, achieving high accuracy often entails increased computational cost, which we strive to minimize in this work.

In some cases, the framework of artificial neural networks (NN) with a special design can also be used to construct an appropriate descriptor of the system \cite{Schutt_quantum-chemical_2017,Schutt_2019, Noe_kinetic_2015, Noe_2019}. 
The most employed framework is Graph Neural Networks (GNNs). 
Initially introduced in 2017 \cite{gcn_2017}, GNN became the most popular DL method for structural analysis in chemistry, biology, and drug design applications. 
Highly-accurate protein structure prediction with AlphaFold \cite{alphafold_2021} has further boosted the interest of different communities in GNN.
Training a GNN consists of two main steps, forming together the propagation rule which is the base of the method: 
(a) aggregating information from neighboring nodes, and (b) updating the nodes and/or edges messages. 
Notably, the aggregation process is permutation invariant.
In this step (a), the input node features are multiplied with the adjacency matrix of the graph and weight matrices, similarly to dense neural networks. 
In step (b), the resulting product is passed through a nonlinear activation function to generate outputs for the subsequent layer.
Based on various details of aggregation and transformation of the passed message, there are a plethora of graph architectures, the most popular being the graph convolutional network (GCN) \cite{gcn_2017} and graph attention method (GAT) \cite{gat_velickovic_2018}.  
The popularity  of the GNN algorithms in materials science, chemistry, and biology is explained by the fact that the descriptor function (feature representation) is learned at the same time as the regression or classification task. 
Numerous methods based on invariant or equivariant descriptions of LAE were proposed since 2017. 
We cite the most popular:  SchNet, a weighted atom-centered symmetry function and the deep tensor neural network \cite{schutt_schnet_2018, schutt_schnetpack_2019}, ALIGNN - Atomistic Line Graph Neural Network \cite{choudhary_ALIGNN_2021}, Deep Graph Library (DGL)-LifeSci \cite{karypis_dgl-lifesci_2021},  DimeNET - Directional Message Passing Neural Network \cite{gasteiger_dimenet_2020, gasteiger_dimenetpp_2020}, PAINN - polarizable atom interaction neural network \cite{schutt_painn_equivariant_2021}, Nequip - Neural Equivariant Interatomic Potentials \cite{batzner_nequip_2022}, DeepMD-kit \cite{WANG2018178}, Allegro – a strictly local, equivariant deep neural network interatomic potential architecture \cite{musaelian_allegro_2023}, mACE - Higher Order Equivariant Message Passing Neural Networks \cite{mace_batatia_2022, mace_design_2022, mace_van2022hyperactive}. 
The top three most accurate force fields in chemical or materials science databases, such as MD17, DB9 \cite{batzner_nequip_2022, musaelian_allegro_2023} and 3BPA \cite{mace_batatia_2022}, are methods based on GNNs. 
GNN-based approaches have demonstrated remarkable performance in accurately predicting and characterizing various chemical and materials properties, making them a prominent choice in the field.

Despite the significant progress made, GNNs are still primarily utilized in the domain of chemistry and biology  
with a numerical cost of \SI{}{\milli\second} per atom per CPU wall time, which is a few orders of magnitude larger than the direct traditional 
descriptors mentioned earlier. 
Consequently, in the classification literature of materials science, traditional descriptors or simpler surrogate models are predominantly used as inputs for classifiers \cite{deringer2019machine, bartok_machine_2017}.

Our challenge is to design a lightweight descriptor that is numerically fast and even lighter than traditional descriptors, while maintaining a high level of accuracy. 
We aim to achieve the best of both worlds by combining the advantages of GNNs and traditional descriptors.

Using radial distances or angles between atoms as descriptors constitutes a valuable, rotation-invariant option. 
The introduction of 2-body or 3-body kernels, as proposed by Glielmo and De Vita \cite{glielmo_efficient_k2b_2018}, allows for the creation of highly numerically efficient descriptors. 
Recent applications in machine learning force fields, such as those showcased in Refs.~\cite{vandermause_flare_2020, xie_flare_2021, byggmastar_knb_HEA_2021, byggmastar_knb_Fe_2022}, demonstrate the robustness of this approach. 
However, the numerical efficiency of this formalism is limited to capturing interactions up to 3-body interactions, which may result in reduced accuracy when dealing with systems in the liquid phase. 
On this observation, we have decided to build our descriptor solely based on interatomic distances, excluding angles or higher-order descriptions. 
However, we aim to retain the many-body aspect of neighborhood geometry.

Once the interatomic distances have been selected to maintain rotation invariance, the issue of permutation symmetry arises. 
Manipulating internal coordinates to achieve a permutation-invariant representation can impact the smoothness and efficiency of the descriptors. 
In general, fundamental operations that preserve inversion symmetry, such as \texttt{sum(neighbors)}, \texttt{max(neighbors)}, and \texttt{min(neighbors)}, are commonly used. 
One potential alternative is to sort the complete list of interatomic distances. 
This operation can be viewed from the GNN perspective as an aggregation around a central atom.
In the present work, we sort all the distances relative to specific central atoms (the detailed construction is presented in §\ref{sec:descriptor_def}).
The concept of sorting distances has been previously utilized in \cite{permutation_genetic_2014} to identify variants of a cluster with a specific number of atoms. 
Similarly, in the context of various molecules, the sorting of 2$-$ and 3$-$body kernels has been employed to address the challenge of permutation invariance \cite{permutation_kbag_cnn_atomic_2018}. 
In this case, the constructed patterns are analyzed and learned using a CNN regressor, resulting in the ``k-bag'' model. 
This model has demonstrated its ability to predict structure-energy relationships with accuracy levels that are comparable to those achieved by \textit{ab initio} methods \cite{permutation_kbag_cnn_atomic_2018}.

The use of sorting distances to achieve permutation invariance has also been proposed in the context of collective variables for free energy sampling. 
In this case, permutation invariant coordinates are obtained by sorting the spectrum of the adjacency matrix of a graph constructed based on the bond network surrounding a specific atom \cite{pietrucci_graph_2011}.  
Furthermore, in the context of polymer chains, a descriptor was developed by considering all inter-monomer distances and sorting them using the spectrum of the corresponding covariance matrix. 
This descriptor was used for classifying the liquid and glassy states in polymers \cite{polymer_data-driven_2022}. 
However, this approach is not applicable in materials science simulations due to the $\mathcal{O}(N^2)$ scaling of the descriptor size.

The upcoming section will demonstrate how the use of sorted radial distances around a central atom can enable the classification of local atomic environments, while addressing the challenge of permutation invariance.

\subsection{High-dimensional encoding of local atomic environments in \descriptor{}s}
\label{sec:descriptor_def}

\begin{figure*}[ht]
    \centering
    \includegraphics[width=\textwidth]{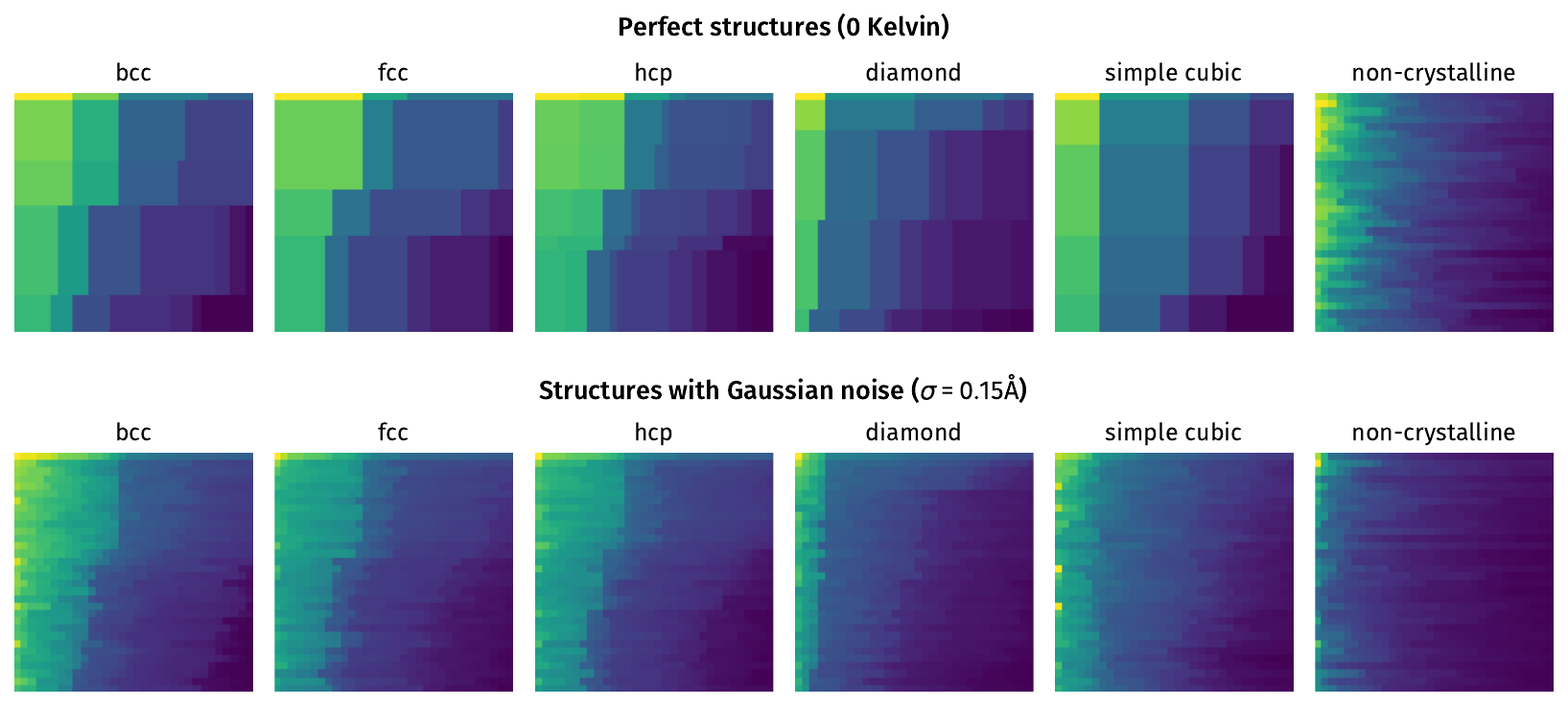}
    \caption{\textbf{Examples of \descriptor{}s for different crystalline and non-crystalline structures.}
    Each image is a \descriptor{} of dimension $32\times32$ with one channel, obtained for an atom in different structures, computed with a value of $r_{cut}$ large enough such that $(n_G-1) \le n(j)$. 
    On the top row, representations of 5 idealized crystalline structures are compared (body-centered cubic, face-centered cubic, hexagonal compact, cubic diamond, simple cubic) and a random non-crystalline system (generated by quasi-Monte Carlo Poisson disk sampling \cite{poisson_disk} in 3 dimensions, for strict illustration purposes), showing differences that can be spotted by the naked eye. 
    In the second row, the same structures were altered by adding Gaussian noise, i.e. displacements $d \sim \mathcal{N}(0, \SI{0.15}{\angstrom})$.
    Corresponding \descriptor{}s show clear differences between the different structures, demonstrating the ability of the method to discriminate crystal structures based on noisy data.
    }
    \label{fig:f2}
\end{figure*}

In this section, we describe our approach to encode atomic environments in high-dimension, using a new descriptor, denoted \descriptor{}. 
The descriptor maps the local neighborhood of a central atom into a 2D image, and provides a pixelated representation of the dense, non-directed graph with weighted edge connections of the neighboring atoms. 
Such image-like representation of atomic structures can be readily combined with a CNN classifier. 
Here, it is worth noting that there have been attempts in the literature to encode atomic environments in a format readable by a standard CNN, as demonstrated in \cite{Fan2021,Fukuya2020}. 
However, these studies encode the raw Cartesian coordinates, and rotational invariance is then achieved through data augmentation. 
In our approach, we aim to directly encode rotational invariance in the descriptor function and bypass the need for data augmentation.

The steps involved in constructing the current descriptor are depicted in Figure 1 and presented in detail below. There are three main steps: (i) Construct the neighbor list within a certain cutoff distance around an atom $j$ and sort it from shortest to longest pair distance.
(ii) Based on the order given by the previous step, create a list of $n_G$ sorted distances from the $n_G$ atoms closest to some central atom of the neighbour list. The central atoms and the closest atoms are within the neighbours list of the $j^{th}$ atom.  
(iii) Using these  $n_G \times n_G$ distances, generate an image of the neighbor graphs. Each central atom is a line in the neighbours map.  
Each step is described in detail in the subsequent three subsections.

\subsubsection{Construction of a fixed-cutoff neighbor graph.} 
\label{sec:descriptor_def1}

We consider a system with $N$ atoms and we denote by $\mathbf{r}_i \in \mathbb{R}^{3} $ the Cartesian coordinates of the $i^{th}$ atom. The Euclidean distance between the atoms $i$ and $j$ is denoted as  $r_{\textrm{ji}} = | \mathbf{r}_{i} -\mathbf{r}_{j}|$. 

To construct the descriptor, we consider the set of neighbors of an atom $j$ within some $r_{\textrm{cut}}$ distance denoted by 
\begin{equation}
    v(j)=\{ i | r_{\textrm{ji}} \le r_{\textrm{cut}}, i \ne j  \}.
\end{equation}
The cardinal of this set $n(j) =  | v(j) |$ is the number of neighbors of the atom $j$ within the $r_{\textrm{cut}}$ distance. 
We then assign labels to the atoms to reflect their spatial proximity to the central atom.
For this, we define a bijective map $\alpha_j : v(j) \to \{ 1, \ldots, n(j) \}$ that transforms the elements of $v(j)$ set into a sequence from $1$ to $n(j)$ such that $r_{j \alpha_j(i_1)}  \le r_{j \alpha_j(i_2)}$  if $\alpha_j(i_1)$ $<$ $\alpha_j(i_2)$.
The map $\alpha_j$ transforms the labels of the neighbors atoms of the $j^{th}$ atom, i.e. $v(j)$ set, into numbers from $1$ to $n(j)$ such that the atom with the mapped number $1$ is the closest to the central atom $j$, 2 is the second-nearest neighbor, and so on until $n(j)$, which is the $n(j)^{th}$ nearest neighbors of the $j^{th}$ atom. 

This allows us to construct a graph $G_j$ having $n(j)+1$ nodes, given by all the atoms from the set $v(j)$ of neighboring atoms and the central atom $j$. 
The nodes of the graph $G_j$ are labeled by $\alpha_j$, from 1 to $n(j)$ with the $0^{th}$ node of the graph corresponding to the atom $j$ itself. 
The edges of the graph $G_j$ are the $n(j)(n(j)+1)/2$ connections between all the nodes of the graph. 

\subsubsection{Distance calculation and node selection.}
\label{sec:descriptor_def2}
Further, we denote by $r_{\textrm{j:lk}}$ the Euclidean distance from the $l^{th}$ neighbor of atom $j$ within $G_j$ to the $k^{th}$ nearest neighbor of atom $l$ within $G_j$. 
In the particular case $r_{\textrm{j:0k}}$, $k$ corresponds to the atom initially labeled as $i$ given by $\alpha_j^{-1}(k) = i$. 
Consequently, $r_{\textrm{j:0k}}$ is equal to $r_{\textrm{ji}}$. 
Similarly, we can calculate the distance $r_{\textrm{j:lk}}$ between the $l^{th}$ neighbor of node $j\equiv 0$ and the $k^{th}$ nearest neighbor of $l$ within $G_j$.

From the set $v(j)$, we select $n_G$ atoms (the first $n_G$ neighbors). 
This number is typically smaller than the average number of atoms $\langle n(j)\rangle$ found within a distance $r_{cut}$ of the atoms of the dataset, resulting in a final image where each pixel encodes one edge of the graph (see the next paragraph), but it is not a restriction.
Below, we consider the two cases. 
In the first case, $n_G \le n(j)$ for any atom $j$, and $n_G$ is chosen to be a power of 2 for a more efficient treatment by the CNN. 
For the cases when  $n_G > n(j)$  the inverse of all sorted distances $r_{j:l\cdot}^{-1}$ and  $r_{j:\cdot k }^{-1}$ for which $l,k > n_G $ will be attributed to a constant value. 
Throughout this study, we will set the constant value to zero. This is the case illustrated in Fig.~1.

\subsubsection{Construction of the 2D Image}
\label{sec:descriptor_def3}

To the graph $G_j$, we will attribute an image of the local atomic environment of the $j^{th}$ atom.  
From the graph $G_j$ and $n_G$ sorted distances $r_{j:l\cdot}$, with  the centers  $l=0, \ldots, n_G-1$, we will build  a matrix $\mathbf{M}_j \in \mathbb{R}^{n_G \times n_G}$. Each line of the matrix is indexed by $l+1 $ = $1, \ldots, n_G$  and is computed  from the line vector $r_{j:l\cdot}$. 
The first row is calculated as:
\begin{equation}
(\mathbf{M}_j)_{1 k} = g_{j:1} \frac{w_{j:0} w_{j:k}}{r_{j:0k}^\gamma} \, , \, k = 1, \ldots, n_G,
\end{equation}
where $w_{j:0}$ and $w_{j:k}$ are weight factors assigned to nodes $0$ and $k$ in $G_j$, respectively. 
These weights can be set to a value, learned or used as hyperparameters. Their purpose is to encode or ``color'' various properties of the atoms, such as their chemical species.
The utility of these weight factors is demonstrated in Sec. \ref{subsec:cuzr} where a multi-components system is investigated.
The hyperparameter $\gamma$, which controls the decay rate of pixels intensity, is set to a value of 2.

Each subsequent row $l+1$ ($1 < l+1 \le n_G$) of the matrix $\mathbf{M}_j$ corresponds to the $l^{th}$ neighbor of node 0 in $G_j$. 
Its terms are calculated as:
\begin{equation}
(\mathbf{M}_j)_{(l+1) k} = g_{j:l+1}  \frac{w_{j:l} w_{j:lk}}{r_{j:l k}^\gamma} \, , \, k = 1, \ldots, n_G, 
\label{eq:matrix} 
\end{equation}
where $w_{j:lk}$ is the weight factor of the $k^{th}$ neighbor of the $l^{th}$ neighbor of node 0 in $G_j$. 
The factor $g_{j:l}$ assigns a weight for the $(l+1)^{th}$ line of the matrix (i.e., the $l$ neighbors of the node 0 of $G_j$), much like ``attention coefficients'' in graph convolutional neural networks with attention mechanisms.
In this study, $g_{j:1}$ is set to 1, while for $n_G \ge l + 1 > 1$, we use $g_{j:l+1} = 1/r_{j:0l}^\beta$, with $\beta=3$. 
However, these values can also be used as hyperparameters or learned through the neural network's loss function, as in the case of graph attention networks.

The treatment of multi-element systems is straightforward. \label{sec:multi-element}
By using weight factors, denoted $w$ in Eq.~\ref{eq:matrix}, we can create multiple channels that capture various properties of each atom species. 
Each channel can be represented by a matrix $\mathbf{M}_j(\mathbf{w})$, computed for a specific set of weight factors $\mathbf{w}$ defined by the user. 
Following this convention, we can introduce $C$ channels, denoted as $\mathbf{M}_j(\mathbf{w} = \mathbf{w}_c)$, where $c$ ranges from 1 to $C$. 
This means that the neighborhood of each atom is described by $C$ image channels, each with dimensions of $n_G \times n_G$.

The first channel is typically a purely geometric channel, where all weight factors $\mathbf{w}$ are set to 1, providing no specific chemical information about the atoms.
The subsequent channels can be assigned weights that are functions of properties such as mass, covalent radius, atomic number, etc. 
The number of channels to be used for a system containing several different elements is an hyperparameter. 
Its optimal value can vary for different databases, and should be determined empirically in representative cases.
In the case of the Cu-Zr alloy studied in this paper (see Sec. \ref{subsec:cuzr}), we set the number of channels $C=2$, and weights are set to:
\begin{align*}
\mathbf{w}_{c=1} = \begin{pmatrix}
w_{\text{Cu}}\\ 
w_{\text{Zr}}
\end{pmatrix}_{c=1} 
= \begin{pmatrix}
1.0\\ 
1.0
\end{pmatrix},
\mathbf{w}_{c=2} = \begin{pmatrix}
w_{\text{Cu}}\\ 
w_{\text{Zr}}
\end{pmatrix}_{c=2} 
= \begin{pmatrix}
1.0\\ 
1.2
\end{pmatrix},
\end{align*}
to create a contrast between the two different elements on the second channel, while they are treated as equivalent in the first channel.

\subsection{Examples of descriptor images}
\label{sec:examples}
In Fig.~\ref{fig:f2}, we provide examples of \descriptor{}s encoding LAEs found in some common crystal structures, namely bcc, fcc, hcp, diamond, simple cubic, and a synthetic non-cristalline system. 
It demonstrates the strong signature of these structures on their \descriptor{}s, with differences that can be readily seen on the images.
In structures with high symmetry (top row of Fig.~\ref{fig:f2}), a number of interatomic distances are equal, resulting in homogeneous regions or bands on the \descriptor{}s. 
Note that the decaying weight $g_{j:l}$ applied on rows has the effect of diminishing the intensity towards the bottom of the image, in order to increase the importance of the top-left pixels, which encode the distances between the nearest neighbors of the central atom.
The effect of Gaussian noise on atoms positions is also illustrated on Fig.~\ref{fig:f2} (bottom row), and unsurprisingly results in noise on \descriptor{}s too.
However, visual comparison between noisy images and their pristine counterpart shows some level of similarity, and differences still appear between noised images based on different structures.
Put together, these examples intuitively explain how our \descriptor{}s can encode accurately LAEs, with some robustness to positions fluctuations that can be exploited by further analysis using a well-trained model.
The choice and design of this model is the subject of the section below.

\subsection{Structural Analysis Model}
\label{sec:cnn}
In this work, we treat structural analysis as a supervised classification task, applied to the high-dimensional representations of the LAEs.
To analyze the features in \descriptor{} images that reflect the crystal structure (see Fig.~\ref{fig:f2}), we employ a Convolutional Neural Network (CNN) \cite{LeNet1} in a classical supervised classification setting.
We use images encoding atomic environments obtained from MD simulations for training, labeled according to the type of crystal structure. 
In order to minimize the computational cost of the analysis, and considering the clear contrast between images seen in Fig.~\ref{fig:f2}, we intentionally give a preference to simple and efficient CNN architecture, consisting of only a few layers and closely resembling the historical LeNet~\cite{LeNet2} or AlexNet~\cite{Krizhevsky2012} networks.
This design choice ensures rapid computation on a single CPU, making the analysis process accessible without specialized parallel computing hardware (i.e. multicore CPUs, GPUs).
However, such platforms can be used transparently for increased throughput, as they are natively supported by popular deep learning libraries.
The technical details of the CNN are presented in detail in Appendix A.

\subsection{Computational efficiency}

Implemented in Fortran as part of the MILADY package, the descriptor achieved a throughput in the order of $10^5$ at/s/CPU in standard conditions as illustrated in Sec.~\ref{sec:examples}.
This result is on par with traditional geometrical methods, such as PTM, while offering a full characterization of the LAE, resulting in considerably more flexibility. 
In the same setting, the typical CNN used in this work for feature extraction, with $\sim 25,000$ parameters, reached a throughput of more than $10^4$ at/s/CPU at inference. 
Further optimization is possible using hardware acceleration, deep learning optimization techniques, alternative architectures, or more traditional machine learning approaches.

\section{Results}
\label{sec:applications}
In this section, we demonstrate the performance of our method on some challenging cases, typically encountered by the community of computational materials science. 
We mainly focus on the cases where none of the traditional approaches has proven fully satisfactory.

\subsection{Phase classification in distorted crystals}

Here, we demonstrate how our workflow can be applied to classify LAEs of crystalline structures that significantly deviate from the ideal structures. 
The accuracy of the classifier is then compared to a-CNA and PTM algorithms. The RMSD parameter for PTM is set to 0.1 in OVITO.

We consider typical structures found in metals: body-centered cubic (bcc), face-centered cubic (fcc), hexagonal compact (hcp), and cubic diamond (cd) structures.
While the identification of these structures in their pristine form can be performed by traditional methods with good results, the task becomes challenging in disturbed systems, especially in the case of high-temperature, noisy data, or when a fraction of the atoms of the structure are missing~\cite{GRANBERG2023111902,boleininger2023microstructure}.
This represents an important barrier to the analysis of experimentally-obtained atomic positions, such as atom probe tomography (APT) data, which can not be treated using traditional methods of LAE analysis.
These data are indeed characterized by an important noise and a detection rate in the order of $50\%$, meaning that half of the atoms of the structure are missing.

\subsubsection{Computational setup and model training}
To assess the accuracy of our method as compared to traditional approaches in the presence of noise and vacancies in the \descriptor{}, we build a synthetic dataset based on MD simulations of bulk bcc Fe~\cite{Ackland2004}, fcc Al~\cite{Mendelev2008}, hcp Zr~\cite{Mendelev2007} and cd Si~\cite{Stillinger1985} using empirical potentials.
Simulations consist of a temperature ramp in the NPT ensemble increasing from \SI{100}{\kelvin} up to $T=\frac{2}{3}T_M$, where $T_M$ is the melting temperature, with a total simulated time of \SI{10}{\pico\second}. 
Snapshots of atomic positions are sampled every \SI{1}{\pico\second} to constitute the database.
Additionally, following Ref.~\cite{Leitherer2021}, $N$ altered copies of each snapshot are added to the database, where a fraction $x_N$ of randomly chosen atoms are removed.
We chose $\{x_1, x_2 \ldots x_6\} = \{0, 0.1, 0.2, \ldots, 0.5\}$, yielding extremely distorted atomic environments compared to low-temperature structures.
Each simulation cell contains 128 atoms and is periodic in all directions. 
With the 4 different classes and 6 different fractions of missing atoms, each sampled 10 times during a simulation, the database contains a total of 30,720 descriptors, each encoding a LAE.
The CNN classifier has 4 classes, corresponding to each of the structures in the database. 
The database is split between a train and validation subset, containing respectively $80\%$ and $20\%$ of the samples.

\begin{figure}[t]
    \centering
    \includegraphics[width=\columnwidth]{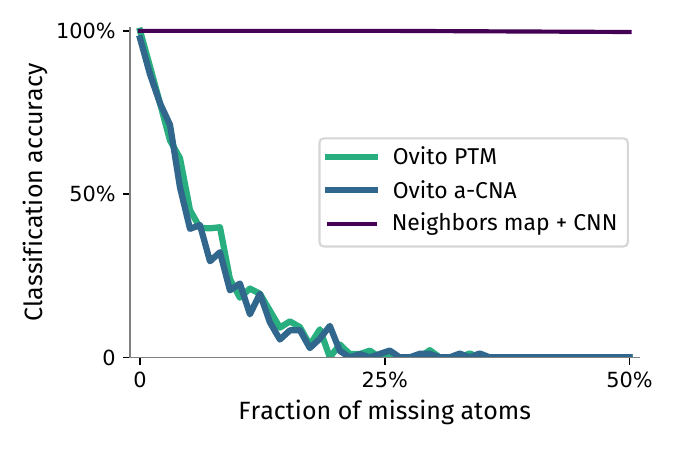}
    \caption{\textbf{Effect of the fraction of missing atoms on accuracy of phase classification.} Overall classification accuracy on a dataset including 4 different phases (bcc, fcc, hcp and cubic diamond). When trained accordingly on structures with missing atoms, the CNN classifier reached near-perfect classification accuracy. Overall accuracy is averaged for all atoms, over the full range of temperature, fractions of missing atoms, and crystalline systems tested.}
    \label{fig:f3}
\end{figure}

\begin{figure}[b!]
    \centering
    \includegraphics[width=0.8\columnwidth]{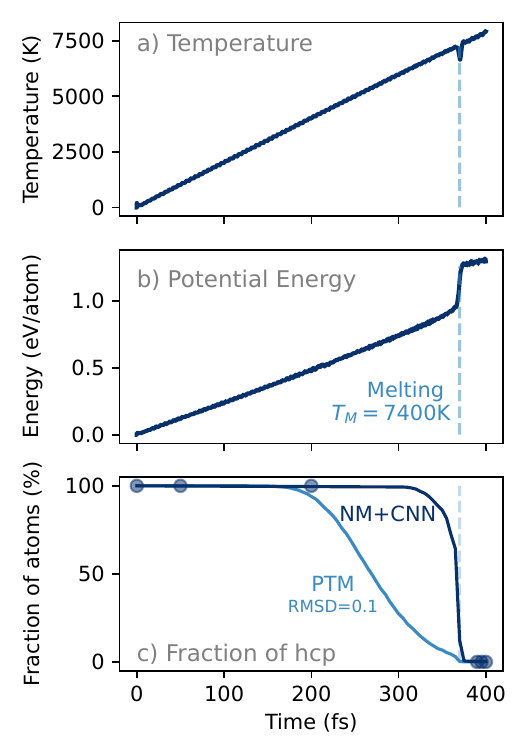}
   \caption{\textbf{Tracking hcp iron structure up to its melting under the Earth's core pressure conditions.} Ramp heating MD simulation at \SI{323}{\giga\pascal}. The onset of melting at \SI{7400}{\kelvin} is indicated with dashed lines in all subplots. (a) Evolution of temperature along the MD trajectory. Temperature is increased up to \SI{8000}{\kelvin} at \SI{2e13}{\kelvin\per\second} rate. (b) Energy curve along the trajectory. (c) Fraction of atoms classified as hcp structure, using PTM (light blue) and our \descriptor{} (NM) + CNN method (dark blue) trained in high-confidence snapshots, indicated with light blue points in (c). 
   }
    \label{fig:f4}
\end{figure}

\subsubsection{Structural analysis}

The overall accuracy of the classifier on the validation dataset is presented in Fig.~\ref{fig:f3}, and compared to the result of traditional methods when applied on the same data.
The accuracy of both a-CNA and PTM are excellent for nearly ideal structures, but they are plummeting when even a small fraction of atoms is missing in the structure.
For the same structures, our method correctly labels highly disturbed atomic environments for the four learned crystal structures, with near-perfect accuracy regardless of the applied noise and missing atoms. Such accuracy indicates that this method can be a promising candidate for further development into a tool for APT data analysis.

Our model demonstrates a satisfying accuracy, comparable with more complex descriptor-based NN classifiers~\cite{Leitherer2021}, but with much better computational performance, thanks to a more efficient descriptor and a reduced number of trainable parameters. 
Training is achieved in a few seconds running on a laptop's CPU, in only 10 epochs and with little hyperparameter tuning --which confirms the computational efficiency of the CNN and its ease of training.
For inference, the model achieved competitive performance on limited CPU resources and, thus, can be used as a drop-in replacement for the a-CNA and PTM methods in any existing data analysis pipeline.

\subsection{Crystal structure identification up to the melting temperature}
\label{subsec:melting_fe}

Identifying crystal structures up to the melting point is a challenging task, as thermal noise can become dominant at temperatures typically starting from $\frac{1}{2}T_M$ to $\frac{2}{3}T_M$, from which standard analysis methods can fail to identify crystalline structures, as we will demonstrate in this section. 
In crystals put under high hydrostatic pressure, the melting point is considerably increased, which can result in a range of a few thousand Kelvins where standard structural analysis is unfeasible.
This is typical of iron when placed in Earth-core conditions, which constitutes an interesting model system to assess the capabilities of our analysis method.
To this end, we simulate the hcp $\to$ liquid phase transformation in a large-scale system using molecular dynamics, based on a recent EAM potential which reproduces well this transformation~\cite{Sun2022}.
As our method is based on supervised learning, one additional challenge for classifying simulated data is the availability of a dataset of annotated LAEs. 
Rather than resorting to a pre-training on synthetic data (which constitutes a valid alternative), we show that traditional structural analysis methods can be used as a source of labelled data. 
As our method can be trained on small datasets, we perform training on a small subset of the simulation data, using appropriate data augmentation, to label the full trajectory.

\subsubsection{MD simulations}

We heat a crystal of hcp iron up to the liquid phase and track the crystal structure evolution. 
To achieve this, we initially relax a crystal of hcp iron containing 108,000 atoms of dimensions $63.8\times 110.5 \times 104.2$~\SI{}{\angstrom}, under a hydrostatic pressure of \SI{323}{\giga\pascal}, to a \SI{1e-3}{\electronvolt\per\angstrom} force tolerance on atoms.
The crystal is then heated from \SI{100}{\kelvin} to \SI{8000}{\kelvin} in the NVT ensemble at a rate \SI{2e13}{\kelvin\per\second}, under the same applied pressure.
The temperature-time curve is reported in Fig.~\ref{fig:f4}a, showing a linear increase. 

In Fig~\ref{fig:f4}~(b), the potential energy curve is reported, showing a continuous increase associated with the temperature ramp, and a sharp step that is typical of first-order phase transitions.
This increase in energy is associated with the collective loss of crystallinity, providing a global indication of when most of the hcp $\to$ liquid transformation occurred.

\subsubsection{Structural analysis}

In Fig~\ref{fig:f4}~(c), we represent the fraction of hcp atoms found in the cell using PTM. 
Note that a-CNA could be used instead of PTM, with similar results.
Starting at about \SI{4000}{\kelvin} or roughly $T_M/2$, the fraction of hcp atoms detected by PTM diminishes and slowly decays to reach 0\% at \SI{7400}{\kelvin}, i.e. when the phase transformation occurs. 
However, the linear evolution of the potential energy between 4000 and \SI{7400}{\kelvin} suggests that no phase transformation occurred in this range, and that the structure remains hcp up to the high temperatures.

To verify this, we set up a CNN classifier to label LAE between the hcp and liquid phase.
We build a database by selecting snapshots of the system taken along the course of the simulation.
In a conservative approach, the snapshots are taken in the domains where PTM accurately labeled the atoms (below \SI{4000}{\kelvin} and above \SI{7400}{\kelvin}), as identified by blue symbols on Fig~\ref{fig:f4} c).
Crystalline environments identified by PTM can be used as a source of conservative (i.e., limiting the risk of mislabelling) annotations to build a training database without user annotation.
While the total amount of annotated samples in the training database can be made arbitrarily large with no additional human effort, we choose to keep its size small to limit training time and demonstrate the method's efficiency.
We thus select 6 snapshots along the trajectory, from which we randomly select only 2,000 atoms from each snapshot (i.e., $\sim2\%$ of the supercell), resulting in a set of 6,000 hcp and 6,000 liquid atoms.
To improve the representation of noised, high-temperature crystals, without using more labelled data, we perform data augmentation by adding a small amount of noise to a fraction of the atoms. 
 For this, we prepare an altered version of each snapshot, where a Gaussian noise of the same order of magnitude as the RMSD of thermal noise, $d \sim \mathcal{N}(0, \SI{0.05}{\angstrom})$ is added to 33\% randomly selected atoms in the snapshot.
Similarly, we draw 2,000 samples from each of these altered snapshots, resulting in a final training dataset of 24,000 structures (the smallest in this paper). 
After training, we use the model to analyze the full trajectory, resulting in the curve shown in Fig.~\ref{fig:f4}~c).
Contrary to PTM, our method yields a much more stable signal, and the phase transformation appears with a strong signature.
The fraction of hcp atoms remains close to 100\% up to \SI{7000}{\kelvin}, where it starts to decrease slowly, and produces a sharp decrease associated with melting, in agreement with the energy curve (dashed vertical line). 

This result is satisfyingly outperforming traditional methods, especially considering the limited database size. 
It demonstrates how the use of traditional methods and data augmentation can help solve the classical challenge of labelled data availability in supervised learning, as zero user annotation was needed.
These results offer an alternative to standard methods for tracking the underlying physics of this phase transition. 
One interesting finding is the intermediate role of the bcc phase, recently spotted \cite{Sun2022}, which is a significant topic in itself but goes beyond the scope of the present investigation.      

\subsection{Melting and vitrification of Ni}

Metallic melts can be solidified into a glassy state through rapid quenching down to low temperature.
In monoatomic metallic systems, Nickel can be used as a model material for the study of metallic glasses ~\cite{Trady2016, Goryaeva_PRM2019}.
However, discriminating between the liquid and amorphous states, which are both non-crystalline phases, remains notoriously challenging. In this section, we examine the ability of our model to distinguish those states in MD simulations.

\subsubsection{MD simulations}
\begin{figure*}
    \centering
    \includegraphics[width=14cm]{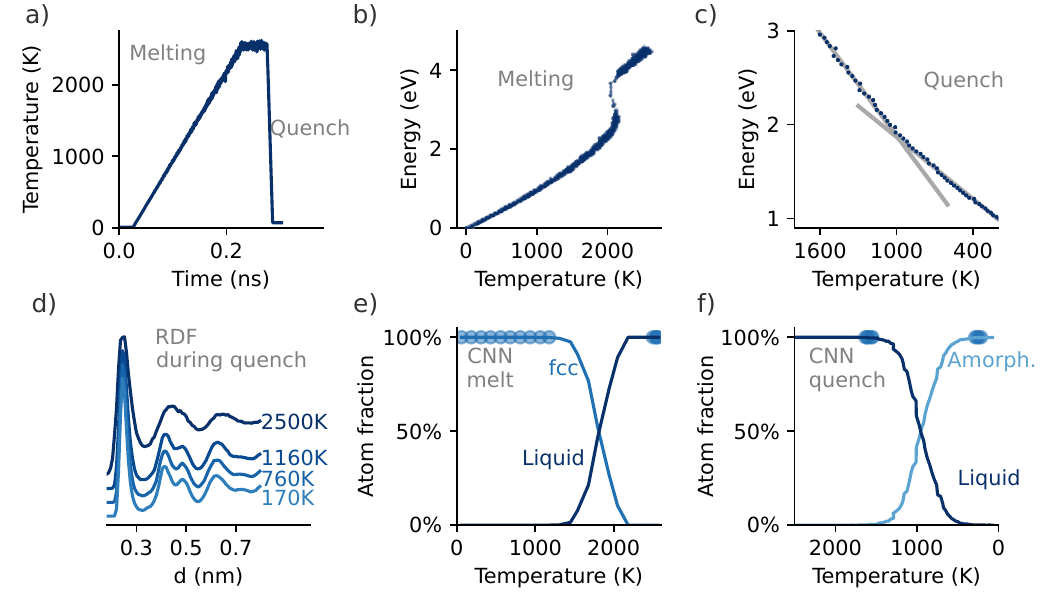}
    \caption{\textbf{Tracking the vitrification of $\mathbf{Ni}$}. (a) Evolution of the temperature during MD simulation. (b-c) Energy as a function of temperature during melting (b) and quench (c), where the inflection point is highlighted. Note the reversed $x$ axis in (c). (d) Partial pair correlation functions obtained at different temperatures along the quench simulation. (e-f) Fraction of different phases identified using our model during melting (e) and quench (f). Configurations present in the database are identified by blue symbols in (e) and (f).}
    \label{fig:f5}
\end{figure*}

The amorphous Ni structure is obtained by the rapid cooling of liquid Ni, following the procedure described in Ref.~\cite{Goryaeva_PRM2019}. 
The simulation cells are fully periodic and contain 6912 atoms (12$\times$12$\times$12 cubic fcc cell). 
In order to achieve vitrification of the structure, the relaxed fcc Ni undergoes the following steps, as shown in Fig.~\ref{fig:f5}~a. First, the temperature is gradually increased up to \SI{2500}{\kelvin} (\SI{1e12}{\kelvin\per\second} rate) within the NPT ensemble in order to melt the system. 
Then, the liquid is equilibrated at \SI{2500}{\kelvin} in the NVT ensemble for \SI{100}{\pico\second}. 
To induce vitrification, the liquid is rapidly quenched (\SI{1e14}{\kelvin\per\second}) down to \SI{10}{\kelvin} within the NPT ensemble.

\begin{figure}[ht]
    \centering
    \includegraphics[width=\columnwidth]{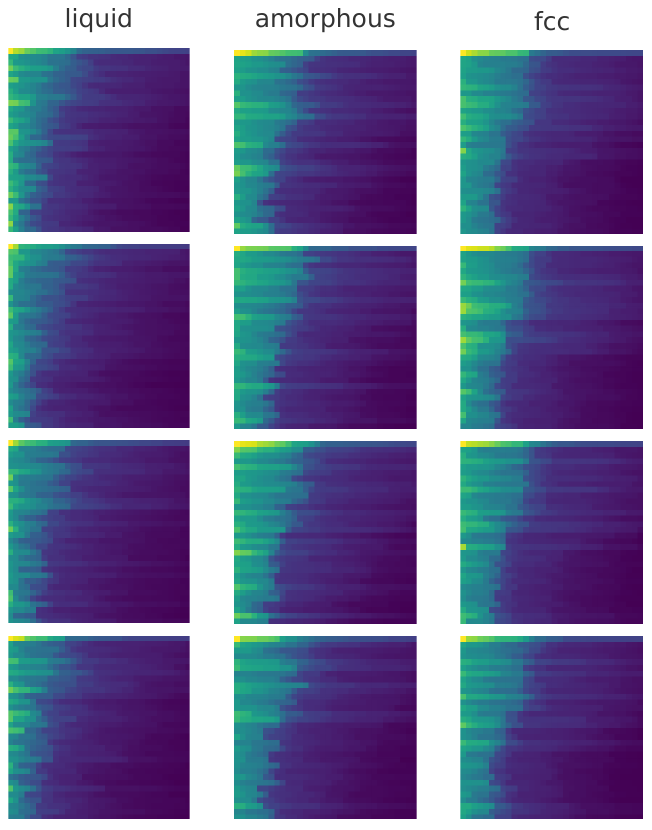}
    \caption{\textbf{Examples of \descriptor{}s from the training dataset of Ni.} Each column corresponds to a different class of structures, from which 4 samples were randomly drawn for illustration purposes.}
    \label{fig:f6b}
\end{figure}

The outcome of the simulations is summarized in Fig.~\ref{fig:f5}. 
During heating up to the melting temperature, the potential energy of the system (Fig.~\ref{fig:f5}b) gradually increases. 
A distinct step of energy increase near \SI{2000}{\kelvin} is caused by the release of the heat of fusion when melting occurs, and serves as a global indicator of the fcc $\to$ liquid first-order transformation.
During a subsequent  quench of the system, the temperature, and potential energy decrease, as shown in Figs.~\ref{fig:f5}a,c.  
A subtle  change of slope (Fig.~\ref{fig:f5}c) near \SI{1000}{\kelvin} indicates a transformation from liquid to a glassy structure. 
This structural transformation can be also distinguished in the global system feature, such as the radial distribution function (RDF) in Fig.~\ref{fig:f5}d, where a splitting of the second peak in the range of 0.4 to \SI{0.5}{\nano\meter} suggests a short-range atomic reorganization into an amorphous structure \cite{Trady2016, Goryaeva_PRM2019}.
The energy and RDF provide global indicators of the transformation of the system, however, they do not give any local, per-atom information. 

\subsubsection{Structural analysis}
To provide a local structural indicator for non-crystalline systems, we use a similar approach as in Sec. \ref{subsec:melting_fe}.
We build a database of structures representing each phase (fcc, liquid, amorphous), which will constitute the three classes of the classification model.
As per-atom manual annotation is practically unfeasible, we label simulation snapshots based on their energy and RDF curves in domains that we consider as high confidence, i.e. below $\frac{2}{3}T_m$ for the fcc structure, from the NVT equilibration at \SI{2500}{\kelvin} and at \SI{1600}{\kelvin} during quench for the liquid, and from the lower temperature domain of the quenching for the amorphous state. %
These snapshots are represented by blue disk symbols in Fig.~\ref{fig:f5}e (for the melting stage), and in Fig.~\ref{fig:f5}f (for the quenching).
A total of 24 supercells are labelled and included in the dataset (10 fcc, 9 liquid and 5 amorphous). 
The lower statistical representation of the amorphous phase is compensated by setting a two times larger weight on this class for loss calculation. 
The database used for training is then composed of a total of 165,888 \descriptor{} representations of LAEs.
A few examples of images, taken from each class of the dataset, are presented in Fig.~\ref{fig:f6b} for visual comparison. 
While inter-class comparison (i.e.~between images of different columns in Fig~\ref{fig:f6b}) may present some similarity, intra-class comparison (i.e.~row-wise) allows identification of some distinctive features, which could be picked up by the model during training.
No data augmentation is performed, to avoid any risk of overlap between the distributions of liquid and glassy structures.

Once trained, the CNN is used in inference to label the LAE of each individual atom of the cell in all frames, resulting in the curves shown in Fig.~\ref{fig:f5}e,f. 
In the domains where structural transformations occur, the model is able to predict a smooth transition from fcc to liquid, capturing with great accuracy the melting temperature (Fig.~\ref{fig:f5}e), and then from liquid to amorphous (Fig.~\ref{fig:f5}f). 
This application demonstrates the accuracy and transferability of our approach, where 864,000 atoms were labelled based on only 24 full-cell user annotations.

\subsection{Vitrification of a Cu-Zr alloy}
\label{subsec:cuzr}

In this section, we showcase the ability of our method to treat multi-component alloys and consider a phase transformation from liquid to amorphous state in Cu-Zr system.

\begin{figure*}[ht]
    \centering
    \includegraphics[width=15.5cm]{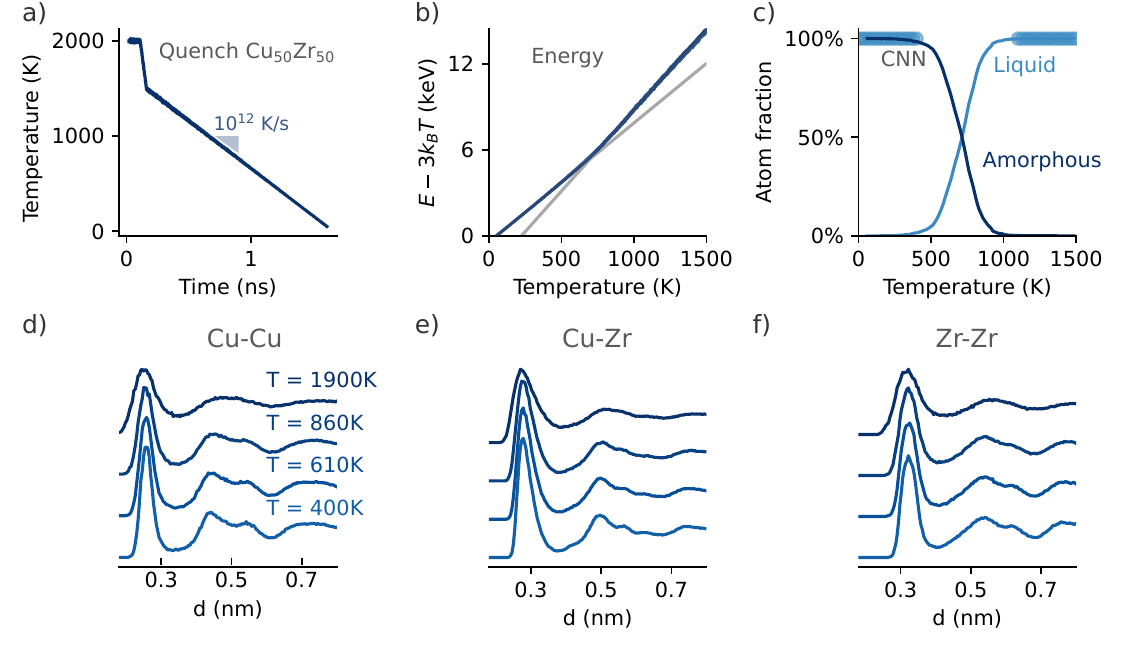}
    \caption{\textbf{Tracking the vitrification of $\mathbf{Cu_{50}Zr_{50}}$}. (a) Evolution of the temperature during quench. (b) Energy as a function of temperature during quench (blue line). Least-squares linear regressions performed before and after the inflection point are plotted as gray lines as an eye guide. (c) Fraction of liquid and amorphous phase detected by our model. Training database configurations are indicated by blue symbols. (d-e-f) Partial pair correlation functions obtained at different temperatures (from dark blue to light blue) along the quench simulation.}
    \label{fig:f6}
\end{figure*}

\subsubsection{MD simulations}

The database preparation consists in creating a liquid structure of the alloy, by thermalizing it in the liquid phase, and subsequently cooling it down to form an amorphous structure.
Initially, we generate the liquid structure by randomly placing 16,000 Cu atoms and 16,000 Zr atoms, resulting in a $\mathrm{Cu}_{50}\mathrm{Zr}_{50}$ composition, in a cubic cell of side \SI{85}{\angstrom}. 
Creating atoms at random positions, there is a risk of atoms being generated too close to one another, leading to an enormous force experienced by the atoms if relaxed with the EAM potential directly, due to its divergence at short distance.
To circumvent this issue, we perform a preliminary minimization step to ensure that all atoms are located further apart than a certain cutoff distance $r_c$. 
For this purpose, we employ the ``soft'' pair style in LAMMPS. 
The pairwise interaction between any two atoms $i$ and $j$, regardless of their species, is given by the potential energy function:
\begin{equation}
E(r_{ij}) = A \left [ 1 + \cos \left ( \frac{\pi r_{ij}}{r_c} \right ) \right ], \quad r_{ij} < r_c,
\end{equation}
with $r_c$ = \SI{1.8}{\angstrom} and a prefactor $A$ = \SI{10}{\electronvolt}. 
The minimization process continues until the force exerted on the atoms falls below \SI{0.1}{\electronvolt\per\angstrom}.
Then, we redefine the soft pairwise interaction with a larger cutoff, $r_c$ = \SI{2.8}{\angstrom}, and a prefactor $A$ = \SI{100}{\electronvolt}, and carry out 10,000 steps of molecular dynamics (MD) in the NVE ensemble, enforcing a maximum displacement of \SI{0.1}{\angstrom} per timestep $t$ = \SI{1}{\femto\second}.
This is done to ensure that most pairwise distances approximate $r_c$.
After these steps, we replace the soft pairwise interaction with the EAM potential from Mendelev et al.~\cite{mendelev2019development}. 
To further stabilize the system, we execute 10 MD steps in the NVE ensemble, maintaining the same maximum displacement limit.
We then carry out thermalization in the NPT ensemble at a constant temperature of \SI{2000}{\kelvin} for 100,000 steps, initializing the system with a Gaussian velocity distribution.
Finally, we conduct two quenching steps, which are descending temperature ramps. 
The first quenching step reduces the temperature from \SI{2000}{\kelvin} to \SI{1500}{\kelvin} at a cooling rate of \SI{1e13}{\kelvin\per\second} to limit the simulated time dedicated to the liquid phase. 
The second quenching step lowers the temperature from \SI{1500}{\kelvin} down to \SI{50}{\kelvin} at a slower cooling rate of \SI{1e12}{\kelvin\per\second} to ensure vitrification.
We collect 5 snapshots taken at intervals of \SI{10}{\pico\second}, and throughout the quenching process, we collect snapshots of the system every \SI{20}{\pico\second} to constitute the database.
In Fig.~\ref{fig:f6} d), the intra- and inter-species RDF are plotted, showing differences between pairwise distances of atoms depending on their species.
To account for the different chemical components, we use a two-channel \descriptor{}, as described in Sec.~\ref{sec:multi-element}, and a 2-channel CNN, with the same architecture and hyperparameters as single-channel CNNs used in other applications.

The results of the MD simulations for the Cu-Zr system are outlined in Fig.~\ref{fig:f6}. 
Similarly to the Ni case, we track the evolution of potential energy during the simulation, identifying the liquid-to-glass transition through an energy shift near \SI{700}{\kelvin} in Fig.~\ref{fig:f6}a. This serves as a global indicator of the transition. 
The modification in the RDF at around \SI{0.5}{\nano\meter} as seen in Fig.~\ref{fig:f6}b, further corroborates the transition to an amorphous structure. 
However, these global indicators lack local, per-atom information and the location of the inflection point on the energy curve in Fig.~\ref{fig:f6}b is challenging to extract precisely based solely on the structural information. 
Furthermore, in liquid and amorphous Cu-Zr phases, traditional LAE analysis methods such as PTM or a-CNA are not applicable, as they are strictly limited to crystalline structure identification.

\subsubsection{Structural analysis}
To perform the structural analysis in Cu-Zr, we proceed as in the Ni case, building a database of structures for each phase, now taking into consideration the liquid and amorphous phases of Cu-Zr, and encoding LAEs in two channels (see Sec.~\ref{sec:multi-element}). 
Simulation snapshots are labeled based on energy and RDF in the high-confidence domains to construct the training set for our model. 
These domains are marked in Fig.~\ref{fig:f6}c, and correspond to less than \SI{450}{\kelvin} for the amorphous, and above \SI{1100}{\kelvin} for the liquid phase.
To limit the size of the dataset while preserving its diversity, $n=10,000$ randomly selected atoms are drawn from each snapshot of the database.
Our database for training then consists of 330,000 LAEs in total (each will be encoded in two channels), with $16\times n =160,000$ liquid and $17\times n=170,000$ amorphous environments.
No data augmentation is used in this case.

Once trained, the CNN delivers individual labels for each atom throughout all frames, revealing the progression from liquid to amorphous phase as indicated by the curves in Fig.~\ref{fig:f6}c. 
In the 400-\SI{1100}{\kelvin} domain, the CNN demonstrates an accurate depiction of the transition at \SI{700}{\kelvin}. 
Thus, our method, similar to the Ni case, showcases its precision and transferability. 

In conclusion, our approach successfully aids in the identification of the liquid-to-amorphous phase transition in a complex, multi-component system. 

\section{Discussion}

The accuracy of the analysis in the applications presented above outperforms traditional methods and competes with methods based on GNNs~\cite{Banik2023} or on descriptors coupled with neural networks \cite{Leitherer2021}, while maintaining much lower computational cost.
Furthermore, the results should be balanced with our decision to prioritize simplicity and computational efficiency, particularly in the design of the CNN that extracts features from descriptors and performs the classification.
In machine learning, classifying real data with a finite set of classes is a common challenge, as some samples might deviate from all known classes. 
The used CNN classifier assigns likelihood scores to each class and selects the most likely one.
Low-confidence samples thus cannot be labeled as ``unknown''. 
It is then unclear if misclassification is a result of limitations of the model, or from samples that significantly deviate from all reference structures and have an undefined environment. 
To handle low confidence predictions in practical applications, an ``unknown'' label could be assigned if the maximum value of the one-hot output vector of the CNN falls below a threshold value.

It is also worth noting that there is no inherent limit to the complexity of the network that can be used to extract features from images, while we intentionally used networks with simple architectures and a few tens of thousands of parameters.
The obtained computational efficiency particularly stands out compared to GNN methods, both in terms of computational load and ease to train the model, needing only tens of thousands of annotated images.
More advanced networks, such as ResNets~\cite{He2016} with millions of parameters, could be applied in applications where the learning capacity of small CNNs would be limiting.
This could happen for extensive training databases where differences between \descriptor{}s are difficult to detect.
This would however necessitate the use of specialized hardware to compensate for the increased computational cost.
For this reason, exploring the performance of more complex models remains out of the scope of the present study.
Similarly, more complex workflows where correlations between \descriptor{}s found in the same system (e.g., for object detection or clustering), and/or between \descriptor{}s obtained in different frames of a trajectory (time series analysis), are analyzed using NNs represent an interesting perspective for the future studies.

Previously, Goryaeva \textit{et al.} \cite{Goryaeva2020} have demonstrated the usefulness of spectral descriptors as embeddings for LAEs, i.e., as vectors encoding a LAE and allowing a direct comparison with other LAEs using an appropriate metric, like Mahalanobis statistical distance.
Following a similar approach with our lightweight descriptor could be beneficial to enable efficient clustering, outlier detection, as well as for learning from very limited amounts of annotated samples (\textit{few-shots} learning). 
This can be addressed with minimal modification of our approach, by learning a similarity metric between \descriptor{}s, e.g., using a Siamese network architecture~\cite{chopra2005learning} to directly optimize the $L_1$ distance between low-dimensional embeddings calculated by the network (see Fig. \ref{fig:f1}), while taking advantage of the ``semantic'' high-level features extraction performed by the CNN.
Storing low-dimensional embeddings in a database can also take several orders of magnitude less space than raw \descriptor{}s and enable faster comparison.
Adaptation of such methods is a promising perspective for further investigations.

Contrary to the descriptors used for interpolating the energy landscape of atomic systems, \descriptor{}s are tailor-made for structural analysis and do not necessarily have to be differentiable.
Therefore, sorting operations can be applied for their construction (see Section \ref{sec:descriptor_def}).
It is worth noting that efficient differentiable sorting is an open problem and some recently proposed solutions~\cite{blondel2020fast} can be considered for further development of \descriptor{}s.

The integration of the present method in HPC environments, for on-the-fly analysis or the treatment of large-scale systems, is straightforward. 
The computation of descriptors in a supercell containing $N$ atoms has a complexity $\mathcal{O}(N)$, and can be parallelized (see Code Availability section). 

\section{Conclusions}
In this work, we propose a simple and efficient approach for encoding local atomic environments into 2D image fingerprints, called \descriptor{}. The descriptor is based on a graph-like architecture with weighted edge connections of neighboring atoms. 
To enable accurate identification of atomic structures, the \descriptor{} descriptor can be readily combined with image processing methods, like CNN classifiers. 
This workflow for structural analysis achieves an accuracy comparable to specialized neural network architectures (e.g., GNN) and spectral descriptors, while maintaining its computational cost comparable to that of traditional geometry-based methods. Structural analysis with \descriptor{} is  accessible with modest computational resources, e.g., without GPU and massive parallelism, while being scalable up to HPC workloads.

The \descriptor{} descriptor intrinsically encodes geometrical invariance, which  allows training on relatively small datasets, in contrast to other NN-based methods for structural identification. Thus, a small subset of the data to analyze --possibly annotated using traditional algorithms in high-confidence regions-- can be sufficient for training the model. 

The proposed approach is applicable to crystalline and non-crystalline structures, including multi-element systems, which are notoriously difficult to interpret both by traditional and recent ML-based methods.
In perspective, the method can be further adapted for the detection and identification of defects, including extended defects and  precipitates in large-scale systems.

The simplicity of the \descriptor{} enables its straightforward implementation across different frameworks, hardware platforms, and programming languages, allowing for easy integration in molecular dynamics engines and post-treatment structural analysis software.
An example of a simple implementation in Python as well as of optimized code suitable for the HPC environment is given in the Code Availability section.

\section*{Acknowledgements}
This work was financially supported by the Cross-Disciplinary Program on Numerical Simulation of CEA, the French Alternative Energies and Atomic Energy Commission. AA, AMG, and MCM acknowledge the support from GENCI - (Jean-Zay/CINES/CCRT) computer centre under Grant No. A0150906973. 
This work has been carried out within the framework of the EUROfusion Consortium, funded by the European Union via the Euratom Research and Training Programme (EUROfusion Grant No. 101052200).
The views and opinions expressed herein do not necessarily reflect those of the European Commission. 
The authors gratefully thank Dr. Isabelle Mouton for the insightful conversations regarding the major challenges in the structural analysis of 3D reconstructions from Atom Probe Tomography.

\section*{ Declaration of competing interest }

The authors declare that they have no known competing financial interests or personal relationships that could have appeared to influence the work reported in this paper.

\section*{Data Availability}
A Python implementation is available on \href{https://github.com/ai-atoms/neighbors-maps}{GitHub (https://github.com/ai-atoms/neighbors-maps)}. 
This implementation was used for generating visualizations in Figs. 1 and 2.
An optimized implementation, available in the \href{https://ai-atoms.github.io/milady-docs/}{Milady package (https://ai-atoms.github.io/milady/)}, was used for the representation of LAEs in section \ref{sec:applications}.

\appendix

\section{Convolutional Neural Network architecture}

\begin{figure*}
    \centering
    \includegraphics[width=1.5\columnwidth]{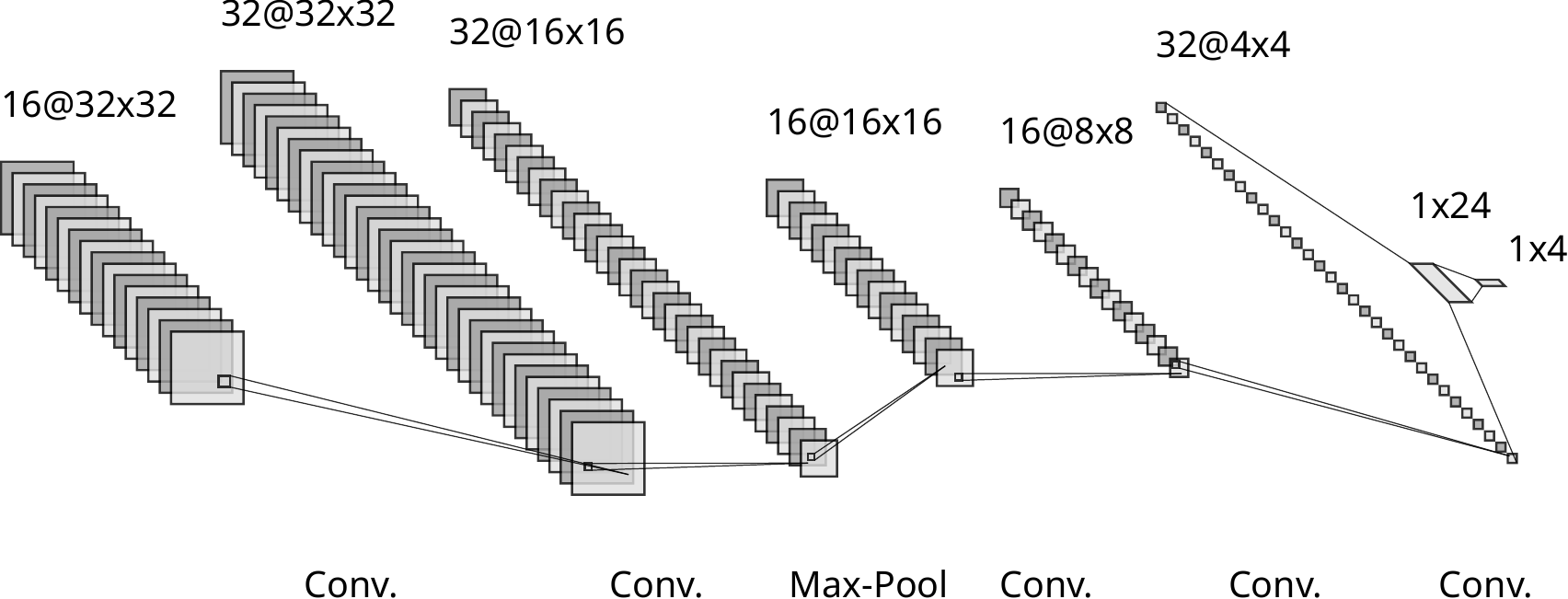}
    \caption{\textbf{Schematic representation of the CNN architecture used in this work.}}
    \label{fig:arch-CNN}
\end{figure*}

The implemented Convolutional Neural Network (CNN) using TensorFlow Keras is sequential with an input shape of (N, 32, 32) for N-channel images. 
The architecture depicted in Fig.~\ref{fig:arch-CNN} consists of:
\begin{enumerate}

    \item A 2D convolutional layer (Conv2D) with 16 filters of kernel size (5, 5) outputting a shape of (16, 32, 32) with 416 parameters.
    \item Another Conv2D layer with 32 filters of kernel size (3, 3), adding 4640 parameters, and maintaining the output shape at (32, 32, 32).
    \item A MaxPooling2D layer with a pool size of (2, 2) to downsample the feature maps to (32, 16, 16).
    \item Two additional Conv2D layers, first with 16 filters (3, 3) and 4624 parameters, reducing the output shape to (16, 16, 16), and then 16 filters with strides of (2,2) and 2320 parameters, yielding an output shape of (16, 8, 8).
    \item A final Conv2D layer with 32 filters, strides of (2,2), 2080 parameters, and a (32, 4, 4) output shape.
    \item A Flatten layer, converting the output to a 1D vector of 512 elements.
    \item A Dropout layer for regularization with a dropout rate of 0.5.
    \item A Dense layer with 24 units using ReLU activation, contributing 12,312 parameters, and a batch normalization layer.
    \item A final Dense layer with 2 nodes for model prediction.
\end{enumerate}

The architecture employs ReLU activations and L2 regularization (1e-4). The model is compiled with the Adam optimizer, Sparse Categorical Cross-entropy loss function, and accuracy as the performance metric, for 10 epochs. 
The model's total trainable parameters amount to 26,490.

\FloatBarrier
\bibliographystyle{elsarticle-num}
\bibliography{biblio}

\end{document}